\documentclass[aps,twocolumn,floatfix,superscriptaddress,amsmath,showpacs,showkeys,prb]{revtex4-1}

\usepackage{t1enc}
\usepackage[final]{graphicx}
\usepackage{graphicx}
\usepackage{epsfig}
\usepackage{bm}
\usepackage[normalem]{ulem}
\usepackage{color}
\usepackage{float}
\usepackage{url}
\usepackage{hyperref}
\usepackage[colorinlistoftodos]{todonotes}

\newcommand{\avrg}[1]{\langle #1\rangle}

\begin{document}
\preprint{APS/123-QED}

\title{A short-ranged memory model with preferential growth}

\author{Ana L. Schaigorodsky}
\email{schaigorodsky@famaf.unc.edu.ar}
\affiliation{Instituto de
F\'{\i}sica Enrique Gaviola  (IFEG-CONICET), Ciudad Universitaria,
5000 C\'ordoba, Argentina}
\affiliation{Facultad de Matem\'atica, Astronom\'{\i}a,
F\'{\i}sica y Computaci\'on, Universidad Nacional de C\'ordoba}

\author{Juan I. Perotti}
\email{juanpool@gmail.com}
\affiliation{Instituto de
F\'{\i}sica Enrique Gaviola  (IFEG-CONICET), Ciudad Universitaria,
5000 C\'ordoba, Argentina}
\affiliation{Facultad de Matem\'atica, Astronom\'{\i}a,
F\'{\i}sica y Computaci\'on, Universidad Nacional de C\'ordoba}

\author{Nahuel Almeira}
\email{nalmeira@famaf.unc.edu.ar}
\affiliation{Instituto de
F\'{\i}sica Enrique Gaviola  (IFEG-CONICET), Ciudad Universitaria,
5000 C\'ordoba, Argentina}
\affiliation{Facultad de Matem\'atica, Astronom\'{\i}a,
F\'{\i}sica y Computaci\'on, Universidad Nacional de C\'ordoba}

\author{Orlando V. Billoni}
\email{billoni@famaf.unc.edu.ar}
\affiliation{Instituto de
F\'{\i}sica Enrique Gaviola  (IFEG-CONICET), Ciudad Universitaria,
5000 C\'ordoba, Argentina}
\affiliation{Facultad de Matem\'atica, Astronom\'{\i}a,
F\'{\i}sica y Computaci\'on, Universidad Nacional de C\'ordoba}

\date{\today}

\begin{abstract}

In this work we introduce a variant of the Yule-Simon model for preferential growth by incorporating 
a finite kernel to model the effects of bounded memory.
We characterize the properties of the model combining analytical arguments with extensive numerical 
simulations.
In particular, we analyze the lifetime and popularity distributions by mapping the model dynamics to 
corresponding Markov chains and branching processes, respectively. These distributions follow power-laws
with well defined exponents that are within the range of the empirical data reported in 
ecologies.
Interestingly, by varying the innovation rate, this simple out-of-equilibrium model exhibits many of 
the characteristics of a continuous phase transition and, around the critical point, it generates time 
series with power-law popularity, lifetime and inter-event time distributions, and non-trivial temporal 
correlations, such as a bursty dynamics in analogy with the activity of solar flares. 
Our results suggest that an appropriate balance between innovation and oblivion rates could provide 
an explanatory framework for many of the properties commonly observed in many complex systems.
\end{abstract}

\maketitle

\section{Introduction} \label{intro}
Usually known as Zipf's law or Pareto distributions, power-laws and long-tailed distributions 
are scattered among many natural systems~\cite{newmann-zipf}. 
For instance, these distributions can be found in physics, biology, earth and planetary sciences, 
economics and finance, and demography and the social sciences.
A short list of the systems where these laws are observed includes the sizes of earthquakes \cite{Bak02PRL-earthquakes} 
and solar flares~\cite{Lu1991-solar-flares}, the frequency of use of words in human languages~\cite{zipf1950}, 
the number of species in biological taxa~\cite{Yule22N-distribution}, people's annual incomes~\cite{pareto1964cours} 
and the distribution of the popularities of game lines in chess~\cite{Blasius09PRL}.

Several mechanisms have been proposed for generating highly skewed  distributions with long tails~\cite{newmann-zipf,Mitzenmacher2003,Corominas-Murtra2015PNAS-zipf-mechanism}.
In particular, the underlying probabilistic mechanism of the Yule-Simon process~\cite{Yule25PTB,simon1955class} 
is one of the most studied due to its simplicity.
The Yule-Simon model, originally inspired by observations of the statistics of biological taxa, reemerged in 
the literature several times. 
In fact, its most recent variant, known as preferential attachment, became one 
of the most important ideas in the early development of the theory of complex networks~\cite{barabasi1999emergence}.
Yule's model was introduced to explain the distribution of the number of species in a genus, family or other taxonomic group. 
Later, a more particular version of the model was introduced by Simon to explain the frequency of words in literary corpuses,
i.e., to explain Zipf's law. 
Within the context of text generation, the Yule-Simon's model works as follows. 
At each discrete time step a new word is generated with probability $p$, or an already existing word is copied with probability $1-p$. 
This process generates a text with few words that are very popular and many others that are barely used.
Namely, the popularity distribution of the words generated by this process is a skewed distribution with a power-law tail of exponent $\alpha = 2 + p/(1-p)$, meaning that the number $n_s$ of words with popularity $s\gg 1$ satisfy $n_s \sim s^{-\alpha}$. 

Another interesting effect that is frequently observed in empirical systems that display 
a long-tailed frequency  distribution, is the presence of memory in the form of long-range correlations\cite{Livina2005,Altmann12PNAS,Schaigorodsky14PA,Montemurro02F} or a bursty 
dynamics~\cite{Altmann09PO,Ishii16PO-memory-texts}. 
These memory effects are not accounted by the standard preferential growth models~\cite{Yule25PTB, simon1955class, Blasius09PRL, Perotti13EPL}.
In order to address the observed memory effects, Cattuto et al.~\cite{Cattuto06EPL} introduced a 
variant to the Yule-Simon model incorporating a fat-tailed memory kernel.
The series generated by Cattuto's model exhibit temporal long-range correlation which are 
missing in the Yule-Simon model~\cite{Cattuto2007, Schaigorodsky16PO}.
Moreover, it preserves the long-tailed popularity distribution  and is able to reproduce 
quite well the statistics of tag occurrences in the collaborative activity of web users~\cite{Cattuto2007}.
In a recent work~\cite{Schaigorodsky16PO} we have shown that the long-range correlations 
introduced by Cattuto's model allow to describe  the memory effects observed in the growing process of an extensive chess database.
However, this model fails to describe the inter-event time distribution in the use of popular game lines 
by individual players. 
In other words, Cattuto's model introduces long-range correlations but fails to produce a {\em bursty} inter-event time dynamics.

Aiming to generate a bursty dynamics, in this work we replace the fat-tailed kernel of Cattuto's model by a finite-sized memory kernel. 
This modification introduces bursts in the occurrence of  sets of words but eliminates the long-range correlations, although its characteristic length exhibits a non-trivial finite size scaling.
It also preserves the power-law distribution expected in Yule-Simon model.
However, the distribution exponent is neither the corresponding for the Yule-Simon model nor Cattuto's model.

The article is organized in the following manner. 
In section I, we introduce the model with the finite-size kernel and analyze the emerging lifetime and popularity distributions. 
In section II we characterize the memory effects, such as correlations and burstiness. 
In section III we perform an analysis of the kernel state as a function of the model parameters to understand the phenomenology of the model,
where we establish an analogy with the statistical mechanics theory of equilibrium system. 
Finally, in Section IV we provide a  discussion of the main results of this work.

\section{Models and popularity distribution analysis}

\subsection{Models}

The mechanism to generate artificial sequences
in preferential growth models is as follows. The process begins
with an initial state of $n_0$ randomly generated words, or more
generally, elements of different classes.
At each discrete temporal step $t$ there are two options:
i) to introduce an element of a new class with probability p or ii) to
copy an already existing element with probability $\bar{p} = 1-p$.
The way to select the element to be copied in option (ii)
establishes the difference between Yule-Simon and Cattuto's
model. In the Yule-Simon model every existing
element has the same probability to be copied, while in Cattuto's
model the probability of copying an element that
occurred at time $t-\Delta t$ is given by the functional form
of the memory kernel $Q(\Delta t)$,
\begin{equation}
Q(\Delta t) = \frac{C(t)}{\Delta t + \kappa_C},
\end{equation}
where $C(t)$ is a  logarithmic normalization factor and $\kappa_C$ is the 
characteristic length or memory kernel extension.

\begin{figure}[H]
  \begin{center}
	\includegraphics[width=0.45\textwidth]{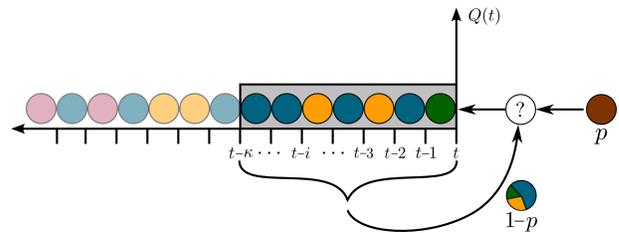}
    \caption{
    \label{resultados-fig-kernel}
    Outline of BMPG model process depicting how an element is incorporated to the time series at time $t$. With
probability $p$ the added element is of a new class, i.e., a new color ball is
introduced. With complementary probability $1-p$ an existing
element is copied, chosen uniformly among the last  elements
of the series. Elements that are more than $\kappa$ time steps away---the shaded balls in the figure---cannot be copied.
}
  \end{center}
\end{figure}

The model we introduce in this work is a simple variant of Cattuto's model. 
The difference resides in the memory kernel, which in our model is defined by a step function:
\begin{equation}
Q(\Delta t) = \left\{ \begin{split}  &\frac{1}{\kappa} \quad & \Delta t\leq \kappa \\ &0 \quad & \Delta t > \kappa \end{split} \right.,
\end{equation}
where $\kappa$ is the memory kernel extension.
We call this variant of the process Bounded Memory Preferential Growth model (BMPG model).
Fig.~\ref{resultados-fig-kernel} outlines the process for the BMPG model.

\subsection{Master equation}

\begin{figure}[H]
  \begin{center}
	\includegraphics[width=0.5\textwidth]{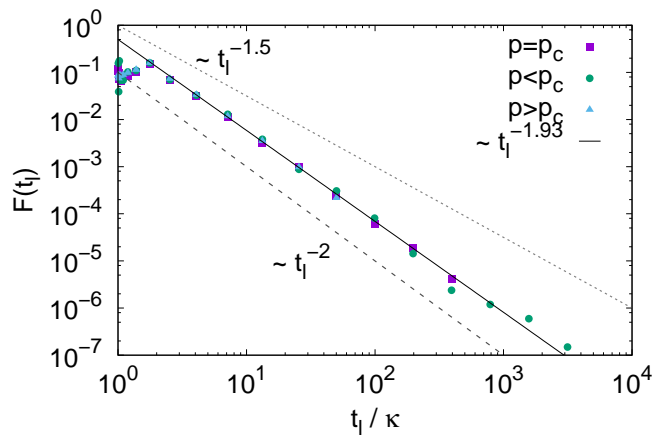}
    \caption{\label{resultados-fig0} Lifetime distribution  calculated for the BMPG model using $\kappa=500$ for
    $p=p_c=1/\kappa$ (violet squares), $p=0.01 p_c$ (green circles) and $p=10 p_c$ (blue triangles). 
    The lines correspond to power-laws with exponents $-1.93$ (full line), $-3/2$ (dotted line) and $-2$ (dashed line).}
  \end{center}
\end{figure}

As a first analysis of our model we describe the dynamics of the elements 
inside the memory kernel. 
Notice that, at each time step $t$, the last element in the kernel leaves 
the kernel in the next time step.
As a consequence, a class of elements may go extinct if the leaving element is the last of its class within 
the kernel---e.g., the pink shaded balls in Fig.~\ref{resultados-fig-kernel}.
In this context, it is interesting to study the extinction dynamics of the elements in the kernel through the analysis 
of lifetime distributions.
Let $P_n(t), \; (n=0,1,...,\kappa)$ be the probability at time $t$ that a given class of element inside the kernel has popularity $n$, and $\textbf{P}(t)$ the vector with components $P_n(t)$  ($n=0,1,2, ... , \kappa$).
The probability $P_n(t)$ can be approached by a discrete Markov chain process, considering that an element of a class can be copied increasing its popularity $n \rightarrow n+1$, can be removed decreasing its popularity, $n \rightarrow n-1$, or can be copied and removed in the same time step keeping its popularity.
In the BMPG process the older element in the kernel is removed. However, to derive the Markov chain we make an approximation, where at each time step we remove instead a randomly chosen element of the kernel. 
In this approximation we derive the following transition probabilities for a kernel with $n$ elements of a particular class,
\begin{eqnarray}
\label{trans-mat}
\Pi_{n,n} &=& (1-p)\left[ \left( \frac{n}{\kappa} \right)^2 + \left( 1-\frac{n}{\kappa} \right)^2 \right] + p\left( 1-\frac{n}{\kappa} \right)(n \neq 0),\nonumber \\
\Pi_{n,n+1} &=& (1-p)\frac{n}{\kappa}\left( 1-\frac{n}{\kappa} \right), \nonumber \\
\Pi_{n,n-1} &=& (1-p)\frac{n}{\kappa}\left( 1-\frac{n}{\kappa} \right) + p \frac{n}{\kappa},
\end{eqnarray}
\noindent where $n=0,1,2, ... , \kappa$. Then, if the probability distribution is known at time $t$, 
 the probability distribution at time $t+1$ can be computed as $\mathbf{P}(t+1) = \mathbf{P}(t)\mathbf{\Pi}$, where $\mathbf{\Pi}$ 
is the matrix with elements $\{\Pi_{n,m}\}$ in which the no null elements correspond to $m=n, n-1, n+1$ and are given
by Eqs.(\ref{trans-mat}).
In particular, using $\dfrac{d\mathbf{P}}{dt} \approx \mathbf{P}(t+1) -\mathbf{P}(t) = \mathbf{P}(t)(\mathbf{\Pi}-\mathbf{1})$  
the master equation can be obtained as
\begin{eqnarray}
\dfrac{d P_n(t)}{dt} &=&  \Pi_{n-1,n} P_{n-1}(t)  + \Pi_{n+1,n}P_{n+1}(t)\\ 
                     & &- (\Pi_{n,n}-1)P_{n}(t), \nonumber
\end{eqnarray}
where the transition matrix is extended to include the special cases $\Pi_{-1,0}=\Pi_{\kappa +1,\kappa}=0$.
Expressing the time in units of $\kappa$---i.e., $\tilde{t}=t/\kappa$---and rearranging terms,
the master equation can be written as:
\begin{eqnarray}
\dfrac{d P_n(\tilde{t})}{d\tilde{t}} &=&  b^{(n-1)}P_{n-1}(\tilde{t})  + d^{(n+1)}P_{n+1}(\tilde{t}) \\
                     & & - (b^{(n)} + d^{(n)})P_{n}(\tilde{t}), \nonumber
\end{eqnarray}
\noindent where $b^{(-1)}=d^{(\kappa+1)}=0$ and, for the other cases,
\begin{eqnarray}
b^{(n)}&=& n \left[(1-p) -(1-p)\frac{n}{\kappa} \right] = n \beta(n) \nonumber  \\
d^{(n)}&=& n \left[ 1 - (1-p) \frac{n}{\kappa} \right] = n(p+\beta(n)).
\end{eqnarray}
\noindent Here $\beta(n) = (1-p)(1 -\frac{n}{\kappa})$.
When $\beta$ is constant this master equation corresponds to a Galton-Watson process~\cite{Pigolotti2005}. 
However, since the number of states is bounded in our model---i.e., the popularity inside the kernel of a 
given element cannot overstep the size of the kernel---$\beta$ is a decreasing function of $n$.
In fitness landscape models, it is usually assumed that the total population size is fixed 
and determined by the carrying capacity of the environment~\cite{Drossel01Review}.
Then, it is interesting to analyze the lifetime distribution for the elements in the kernel, since
former studies used this kind of approach to model the lifetime distribution of species in ecologies~\cite{Pigolotti2005}.
As we will later show, different regimes arise in the BMPG model dynamics depending on the value of the product $p\kappa$.
In Fig~\ref{resultados-fig0} we show the lifetime distributions for the elements in the BMPG model for three different
regimes $p>p_c$, $p=p_c$, and $p<p_c$, where $p_c=1/\kappa$. These distributions, like all the results shown in this paper, 
are obtained by generating sequences of $N=10^7$ elements. 
In these series we compute the lifetime of each class of elements $g$, averaging over $20$ realizations for each
value of the parameters.
In the three cases the lifetime of the elements is distributed according a long-tailed power-law distribution with 
exponent $\approx -1.9$. This exponent is between $-3/2$, which is
the exit time of a random walk problem, and $-2$, the critical Galton-Watson branching process~\cite{Pigolotti2005}. 
Furthermore, it has been suggested~\cite{Sneppen1995} that the lifetimes of taxa in the fossil record 
have a highly right-skewed distribution with a power-law tail with exponents in the range from $-3/2$ to $-2$. 
\noindent

\subsection{Popularity distribution}

\begin{figure}[H]
  \begin{center}
	\includegraphics[width=0.45\textwidth]{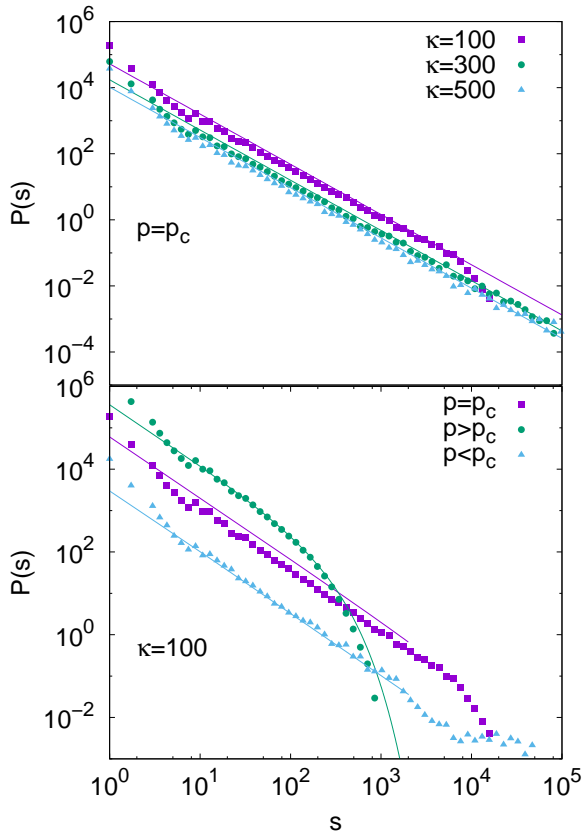}
    \caption{\label{resultados-fig1}(Color online) Top panel: Popularity distribution calculated for 
    the generated sequences with BMPG model for $\kappa= 100$, $300$ and $500$, and $p=p_c$. 
    The straight lines correspond to fits of Eq.~\eqref{eq-Pk}. In all cases the fitted exponents are  
    $\alpha \simeq 3/2$. 
    Bottom panel: Popularity distribution calculated for the generated sequences with the BMPG model for $\kappa= 100$ and 
    $p=0.1 p_c$, $p_c$ and $10 p_c$; the full lines correspond to the estimation in Eq.~\eqref{eq-Pk-4}.}
  \end{center}
\end{figure}

As a further characterization of the model we calculate the popularity distribution $P(s)$ of the generated 
sequences for the BMPG model using various values of the parameters $p$ and $\kappa$.
In Fig.~\ref{resultados-fig1} (top panel) we show the results obtained for $\kappa = 100$, 
$300$ and $500$, where in each case $p=p_c=1/\kappa$. For all calculated $P(s)$ we 
find that the distributions are very well fitted by a power-law,
\begin{equation}
\label{eq-Pk}
P(s) \sim s^{-\alpha},
\end{equation}
with $\alpha \simeq 3/2$, which means that the finite-sized kernel maintains the power-law decay 
found in the original Cattuto and Yule-Simon models, but it does modify the exponents $\alpha$ in a 
particular form as we will discuss below.

In order to analyze the popularity distribution of elements in the BMPG model, we map the generation of elements of each 
class to a corresponding branching process. 
A branching process can be associated to a rooted tree. 
The process begins with a root node which creates children, and those children create children of their own and so on.
The number of children that each node generates is represented by a stochastic variable.
Moreover, each node belongs to a specific generation defined by its distance to the root node of the corresponding tree.
In the scheme of the branching process, the generation of elements of a given class in the BMPG model is approximated 
by the evolution of a corresponding branching process.
Each copy of an element of a given class corresponds to one of the nodes of a rooted tree, except for
the root which corresponds to the first apparition of this class of elements.
In this way, each time the root is copied, a child of the first generation  is born.
In a similar manner, if a first generation node of this element is copied, a second generation child is born and so on.

Each element can have at most $\kappa$ children, as it can only be copied while it is inside the memory kernel.
In this process, the probability that a given node has $i$ children can be well approximated 
by a binomial distribution,
\begin{equation}
\label{eq-Pk-1}
p_i = \binom{\kappa}{i}\theta^{i}(1-\theta)^{\kappa-i},
\end{equation}
\noindent where $\theta = \frac{1}{\kappa}(1-p)$ is the probability that a given element
in the kernel is copied at every time step. In a branching process ruled by a binomial
distribution, each link in the corresponding tree is generated with probability  $\theta$.
Since the number of links in a tree is $s-1$, where $s$ is the number of nodes or elements,
then the probability of generating $s-1$ links is $\theta^{s-1}$. The number of missing links
to complete a $\kappa$-ary tree with $s$ internal nodes is $\kappa s -(s-1)$,  where internal 
means nodes that are not leaves.
Then the probability that this process generates one particular $\kappa$-ary tree with $s$ internal nodes 
and  $\kappa s -(s-1)$ missing links is~\cite{Corral13Book} $\theta^{s-1}(1-\theta)^{(\kappa -1)s +1}$. 
The number of $\kappa$-arys trees with $s$ internal nodes is~\cite{drmota2004combinatorics},
\begin{equation}
\label{eq-Pk-2}
N^{(\kappa)}(s)=\frac{1}{(\kappa -1)s+1} \binom{\kappa s}{s},
\end{equation}

\noindent and then the probability of having a $\kappa$-ary tree with $s$ internal nodes 
and  $\kappa s -(s-1)$ missing links, or in the context of the BMPG model an element
of popularity $s$, is:
\begin{equation}
\label{eq-Pk-3}
P(s)=\frac{1}{(\kappa -1)s+1} \binom{\kappa s}{s}\theta^{s-1}(1-\theta)^{(\kappa -1)s +1}.
\end{equation}
Using Stirling's approximation for $\kappa \gg 1$ and $s \gg 1$ we can obtain 
the asymptotic behavior of Eq.~\eqref{eq-Pk-3},
\begin{equation}
\label{eq-Pk-4}
P(s) \sim  e^{-s/s_0(\kappa,\theta)} s^{-3/2},
\end{equation}
with $s_0(\kappa,\theta) \approx \frac{\kappa -1}{\kappa} \frac{2}{(1-\kappa \theta)^2} = \frac{\kappa -1}{\kappa} \frac{2}{p^2} $.

In Fig.~\ref{resultados-fig1} (bottom panel) we show the resulting distributions 
$P(s)$ calculated from the generated database with the BMPG model, together with 
the theoretical prediction Eq.~\eqref{eq-Pk-4}. It can be seen 
that the theoretical prediction fits very well the distribution calculated from the sequences 
generated by the model, when the values of $\kappa$ are large enough; this confirms that a branching 
process with a binomial distribution is a good approximation in this scenario. 
Moreover, when $p=p_c=1/{\kappa}$,  the exponential cutoff of distribution $P(s)$ (Eq. (\ref{eq-Pk-4})) 
has the form $s_0 \propto \kappa^2$. Therefore, for large values of $\kappa$, such as the ones 
used in Fig.~\ref{resultados-fig1} (top panel), the distribution tends to a power-law, $P(s) \sim s^{-3/2}$, 
in a wide range of values of $s$ as observed in the figure. 

Finally, we fitted the exponents of the frequency distribution for different values of the memory kernel extension $\kappa$ and the probability
of introducing a new element in the kernel (not shown) and we found that for small values of $p$ all 
the exponents $\alpha$ are nearly constant, while for larger values of $p$ they grow as $p$ 
increases. As expected, for $p \leq 1/\kappa$, all exponents have nearly 
the same value  $\alpha \approx 3/2$. 

\section{Memory effects}

\subsection{Correlations}

\begin{figure}[H]
  \begin{center}
	\includegraphics[width=0.45\textwidth]{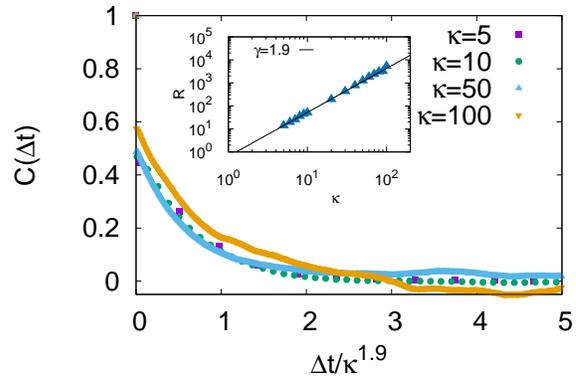}
    \caption{\label{resultados-fig6} Autocorrelation function $C(\Delta t)$ calculated for sequences generated 
    with the BMPG model for $\kappa= 5, 10, 50, \, \text{and} \,100$ and $p=p_c$, rescaled 
    by $\kappa^{\gamma}$ with $\gamma = 1.9$. Inset: Decay characteristic time $R$ of the autocorrelation function 
    as a function of the memory kernel extension $\kappa$. The straight line 
    corresponds to a linear function fit in log-log with exponent $\gamma=1.9$.}
  \end{center}
\end{figure}

As we mentioned in the introduction, the sequences generated in the original Cattuto's model have long-range correlations. 
Since the BMPG model has a finite sized memory kernel, we do not expect long-range correlations for the generated time series $x[t]$.
To build the time series we assigned to each class of elements a random generated number from a Gaussian distribution\cite{Schaigorodsky16PO}. 
In order to see how the correlation length behaves in this kernel, we computed the autocorrelation function 
$C(\Delta t, t)$ of  the time series $x[t]$,
\begin{equation}
\label{eq-correlation}
C(\Delta t,t) = \frac{\langle (x[t]-\mu(t))(x[t+\Delta t]-\mu(t+\Delta t)) \rangle}{\sigma(t)\sigma(t+\Delta t)} ,
\end{equation}
where $\mu(t) = \langle x[t] \rangle$ and $\sigma^2(t) = \langle (x[t]-\mu)^2 \rangle$
and $\langle ... \rangle$ means the expectation value.
Considering the time series is stationary, then $\mu(t)= \mu(t + \Delta t) = \mu_0$,  
$\sigma(t) = \sigma(t + \Delta t) = \sigma_0$ and the autocorrelation function $C(\Delta t,t) = C(\Delta t)$, which can be computed as 
\begin{equation}
\label{eq-correlations}
C(\Delta t) = \frac{1}{N-\Delta t}\sum_{i=1}^{N-\Delta t}\hat{x}[t_i]\hat{x}[t_i + \Delta t] ,
\end{equation}
where $\hat{x}[t_i] = (x[t_i] - \mu_0)/\sigma_0$, $\mu_0 = \frac{1}{N} \sum_{i}^{N} x[t_i]$ and 
$\sigma_0^2 = \frac{1}{N} \sum_{i}^{N} (x[t_i]-\mu_0)^2$. We found that $C(\Delta t)$ decays exponentially,
$C(\Delta t) \sim e^{\frac{-\Delta t}{R}}$, as can be seen in Fig.~\ref{resultados-fig6}.
We calculated the autocorrelation function for several values of $\kappa$ and $p=p_c=1/\kappa$ and 
computed the correlation length R by fitting the exponential decay function. 
In the inset of Fig.~\ref{resultados-fig6} 
we show the correlation length $R$ as a function of $\kappa$. As can be seen $R$ grows 
as a power-law with exponent $\gamma = 1.9$. Fig.~\ref{resultados-fig6} also shows the collapse of the autocorrelation curves when the time is measured in units of the correlation length $R \sim \kappa^{1.9}$.
\subsection{Inter-event time analysis}

Inter-event time analysis is common to many natural phenomena that exhibit memory 
such as earthquakes \cite{Bak02PRL-earthquakes,Corral-PRL-2004},
sunspots \cite{Wheatland98AJ}, neuronal activity \cite{Keat01N} and in several human activities~\cite{barabasi2005origin,jo2015correlated,Schaigorodsky16PO}.
In particular, following the analogy we already established between the occurrence of words in texts and our temporal series, the inter-event time distribution can be analyzed in a similar manner than in texts~\cite{Altmann09PO}.
All the elements in the  sequence have a corresponding time of occurrence. 
Given an element of class $g$ of the time series, $t^{(g)}(j) \in \{1,2,...,N\}$ corresponds to the time of the $j$-th occurrence of $g$. Accordingly, the  $j$-th inter-event time of a particular element $g$ is defined as:
\begin{equation}
\tau_j^{(g)} = t^{(g)}(j+1) - t^{(g)}(j),
\end{equation}
If $s^{(g)}$ is the total number of occurrences of $g$ in the generated series, i.e., its popularity, 
then we can estimate the average  inter-event time as $\avrg{\tau^{(g)}}\approx t_l^{(g)}/s^{(g)}$, where 
$t_l^{(g)}$ is the lifetime of the elements of class $g$,
which is usually called the Zipf's wavelength of word $g$ in  text analysis~\cite{Altmann09PO}.
For simplicity, let us write $\tau$ instead of ${\tau^{(g)}}$ and also introduce
the probability density $f(\tau)$ of inter-event times for any particular class of elements 
in the series.

A bursty dynamics is characterized by a sequence of short periods of high activity followed by long periods 
of low activity~\cite{Goh08EPL}. 
The presence of burstiness in a time series can be studied through the probability distribution of inter-event 
times $f(\tau)$, where any departure of $f(\tau)$ from an exponential distribution may indicate the existence 
of memory in the inter-event time process.
For instance, in the case of words in a text a deviation is observed, which is usually well described by the  
stretched exponential distribution, or Weibull function \cite{Altmann09PO},
\begin{equation}
\label{eq5}
f(\tau)= \frac{\beta}{\tau_0} \left(\frac{\tau}{\tau_0}\right)^{\beta-1} e^{-(\frac{\tau}{\tau_0})^\beta}.
\end{equation}
\noindent For this distribution $\langle \tau \rangle = \tau_0 \Gamma(\frac{\beta +1}{\beta})$,
where $\Gamma$ is the Gamma function and $0<\beta \leq 1$.
If $\beta=1$, burstiness is not present, as is the case for the generated 
series with Cattuto's or the Yule-Simon's models\cite{Schaigorodsky16PO}. 
On the other hand, $0 < \beta < 1$
is an indicator of the presence of burstiness
and, as the value of $\beta$ approaches zero, the apparition of bursts in the time series increases.
A more clear indication of burstiness comes when the distribution of inter-event times  is described by a power-law, 
\begin{equation}
\label{eq6}
f(\tau) = A \tau^{-\eta}, 
\end{equation}
where the degree of burstiness is characterized by the exponent $\eta$ of the power-law; the
lower the exponent, the more bursty the dynamics is. 

Finally, the presence of burstiness in empirical systems can also be characterized by calculating the coefficient of 
variation $\sigma_{\tau}/ \langle \tau \rangle$, where $\sigma_{\tau}$ is the standard deviation of the inter-event times. 
Using this coefficient we compute the burstiness parameter $B$ as~\cite{Goh08EPL}, 
\begin{equation}
\label{eq7}
B = \frac{(\sigma_{\tau} / \langle \tau \rangle-1)}{(\sigma_{\tau} / \langle \tau \rangle+1)}= \frac{\sigma_{\tau}  - \langle \tau \rangle}{ \sigma_{\tau} + \langle \tau \rangle}.
\end{equation}
This parameter is greater than zero for a bursty dynamics and lower than zero when the dynamics becomes regular. 
When $B=0$ there is neither burstiness nor regularity like, for instance, in a Poisson process.

Using the approaches described above, we analyze the inter-event time dynamics, i.e. burstiness, 
of  several elements in series generated by the BMPG model. 
First we define the level of activity of a given element $g$ as the inverse of the Zipf's wavelength,   
as $a_g=1/\langle \tau^{(g)} \rangle$ where  $\langle \tau^{(g)} \rangle$ is the mean inter-event time 
of this element.
Then, analyzing the time series we found that, on average, the activity of the elements depends on their popularity 
through the following relation $a \sim s^{1/2}$, i.e., the most active elements are also the most popular.
Secondly, we analyze the inter-event time distributions for different sets of class of elements. Each set is built  
by aggregating elements with decreasing order of popularity until reaching a specific number of total 
inter-event times. In this way we construct four sets of  $\sim 2 \times 10^6$ inter-event times with
decreasing level of activity.  

\begin{figure}[H]
  \begin{center}
	\includegraphics[width=0.4\textwidth]{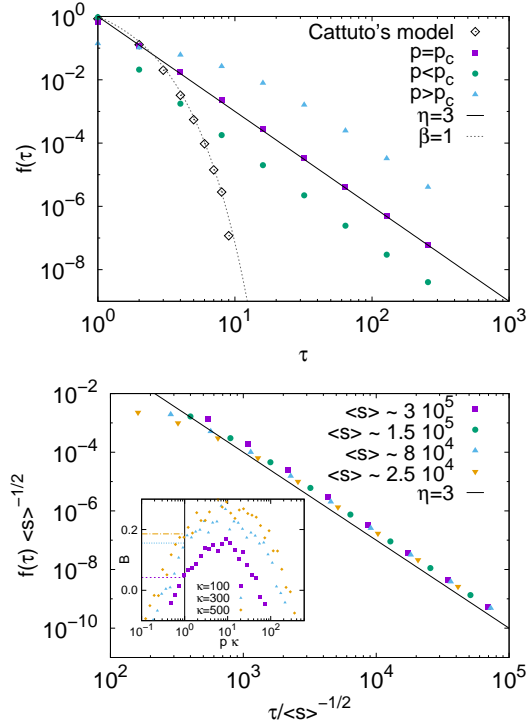}
    \caption{\label{resultados-fig2} Top panel: Distribution of inter-event times measured in generated 
    sequences with the BMPG model for $\kappa=500$ and three values of p; $p=p_c$ (violet squares), $p=0.1 p_c$ (green circles) 
    and $p=10 p_c$ (light blue triangles). 
    We also show a fit according to Eq.~\eqref{eq6} 
    (black full line) for the case of $p=p_c$. 
    For comparison we show the distribution of inter-event times for a generated sequence with Cattuto's model 
    for $\kappa_C=500$ and $p=1/\kappa_C$ (black empty diamonds) and a fit with Eq.~\eqref{eq5} (grey dashed line). 
    Bottom panel: Distribution of inter-event times for different levels of activity, rescaled 
    with $\avrg{s}^{-1/2}$ and a power-law with exponent $3$ for comparison. 
    Inset: Burstiness parameter $B$ calculated 
    for the BMPG model for various values of $\kappa$ and $p$ as function of $p \kappa$; the black vertical line corresponds 
    to $p \kappa=1$ and  the horizontal lines correspond to the values of $B$ calculated from the distribution $f(\tau)$.}
  \end{center}
\end{figure}

In the top panel of Fig.~\ref{resultados-fig2} we show the distribution of inter-event times $f(\tau)$ for the 
most active set,  using $\kappa = 500$ and three different values of the parameter $p$ ($p=0.1 p_c$, $p = p_c$, and $p=10 p_c$). 
As can be seen, $f(\tau)$ has a power-law tail in the three cases with an exponent $\eta = 3$, 
which indicates a clear presence of burstiness. 
For $p=p_c$  the whole distribution is well described by a power-law. 
However, the deviation of  $p$ from $p_c$ affects the distribution for short times.
At $p=0.1 p_c$ the series is homogeneous, i.e. composed almost entirely of elements of a single class, increasing the number of shorter inter-event times.
As $p$ increases ($p = 10p_c$), the number of different classes of elements also increases and the inter-event times are longer, 
resulting in a flat distribution for lower values of $\tau$, as can be seen in this figure.
We also observed that, independently  of the value of $\kappa$ used  in the model (not shown), the decay exponent remains the same.
For comparison, in this figure we also show the inter-event time distribution obtained with Cattuto's model for $\kappa_C=500$ and $p=p_c$ and the corresponding fit using a Weibull distribution with $\beta = 1$.
As we mentioned, in this case the inter-event time distribution is well explained by a Poisson process. 

In the bottom panel of Fig.~\ref{resultados-fig2} we show the distributions of inter-event times $f(\tau)$ for $\kappa=500$ and 
$p=p_c$, for several classes of elements grouped into subsets corresponding to different levels of average popularity.
All distributions present a power-law decay with exponent $\eta=3$ 
and collapse when the axes are rescaled by the factor $\avrg{\tau} = \avrg{s}^{-1/2}$, where $\avrg{s}$ is the average of the popularity of the corresponding subset.

We also calculated the burstiness parameter $B$ for different values of $\kappa$ and $p$, using in this case the 
most active set; see the inset of Fig.~\ref{resultados-fig2} (bottom panel).
For all $\kappa$ there is a range around $\kappa p \sim 10$ where 
a bursty dynamics is evidenced, while for low values of $\kappa p$,  $B<0$, since the series becomes more regular.
As the distribution of inter-event times for $p=p_c$ is a power-law with exponent $\eta=3$, we can calculate 
the corresponding values of $B$ for different values of $\kappa$ by computing $\avrg{\tau}$ and $\avrg{\tau^2}$ directly from the 
normalized distribution, resulting in
\begin{equation}
\avrg{\tau} = \int_{1}^{\kappa} \tau f(\tau) d\tau = \frac{2 \kappa}{\kappa +1},
\end{equation}
\noindent and
\begin{equation}
\avrg{\tau^2} = \int_{1}^{\kappa} \tau^2 f(\tau) d\tau = \frac{2 \kappa^2 \ln(\kappa)}{\kappa^2 -1}.
\end{equation}
\noindent With these expressions we computed  $B$ (Eq.~\eqref{eq7}) for different values of $\kappa$.  
The results are shown in the inset of Fig.~\ref{resultados-fig2} (bottom panel) as horizontal lines. 
As can be seen, the computed values of $B$ coincide with the calculated values from the generated series with 
the BMPG model at $p=p_c$.

\section{Kernel analysis}

\begin{figure}[H]
  \begin{center}
	\includegraphics[width=0.4\textwidth]{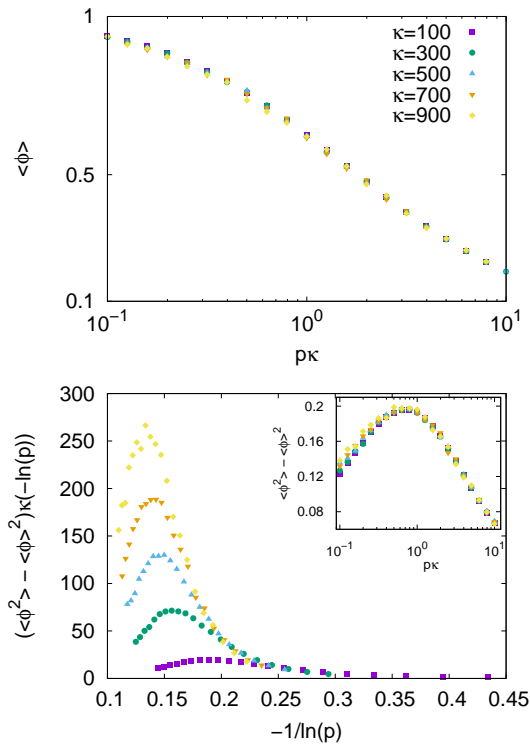}
    \caption{\label{resultados-fig5} Order parameter mean $\avrg{\phi}$ (top panel) and fluctuations $(\avrg{\phi^2}-\avrg{\phi}^2)\kappa/p$ (bottom panel) calculated for the BMPG model for various values of $\kappa$ as a function of $p$ and rescaled by $p_c=1/\kappa$ (Inset).}
  \end{center}
\end{figure}

In order to understand the emergence of burstiness in finite-sized kernels we analyzed the state 
inside the kernel in the BMPG model as a function of the model's parameters.
In analogy with phase transitions, in a first approach we analyzed the state inside the kernel by 
defining an order parameter $0 \leq \phi \leq 1$ as follows, 
\begin{equation}
\label{res-eq3}
	\phi=\frac{1}{\kappa} \max_g n_{\kappa}^{(g)},
\end{equation}
where $n_{\kappa}^{(g)}$ is the popularity of the class of elements $g$ inside the kernel of length $\kappa$. 
According to this definition, when the kernel is full of  elements of a single class, the order parameter is
equal to one.
On the other hand,  when all elements in the kernel are of different classes $\phi = 1/\kappa$, taking its minimum value.
In Fig.~\ref{resultados-fig5} (top panel) we show the mean value of the order parameter as a function of $p\kappa$ 
for several values of the kernel extension $\kappa$, where $\phi$ is computed inside $N/\kappa$ disjoint segments of 
length $\kappa$ all across the generated series ($N$ being the length of the series), and the mean $\avrg{\phi}$ is calculated over 
all $N/\kappa$ segments.
One can note that all the curves collapse when plotted as a function of $\kappa p$, also it can be observed a continuous 
transition from a disordered state  $\phi \sim 1/\kappa \approx 0$ to an ordered state $\phi \approx 1$ as $\kappa p$ decreases. 
The inflexion of the curves that occurs at $p\kappa \approx 1$, suggests of a transition at this point.

\begin{figure}[H]
  \begin{center}
	\includegraphics[width=0.45\textwidth]{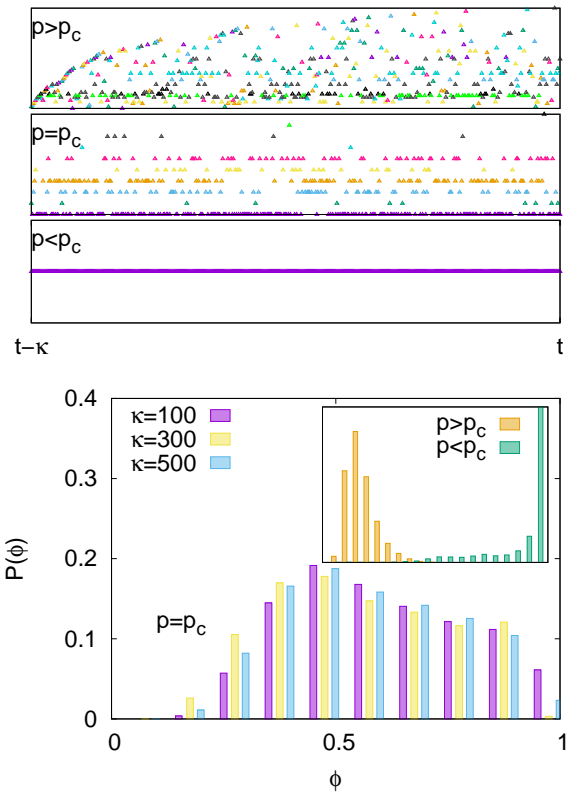}
    \caption{\label{resultados-fig4}(Color online) Top panel: kernel configurations for the super-critical 
    ($p > p_c$), critical ($p = p_c$) and sub-critical ($p<p_c$) states. Bottom panel: Order parameter distribution 
    calculated for the BMPG model for various values of $\kappa$ in the case of $p=p_c$. Inset: Examples of the order 
    parameter distribution in the case of $p>p_c$ and $p<p_c$.}
  \end{center}
\end{figure}

\begin{figure}[H]
  \begin{center}
	\includegraphics[width=0.4\textwidth]{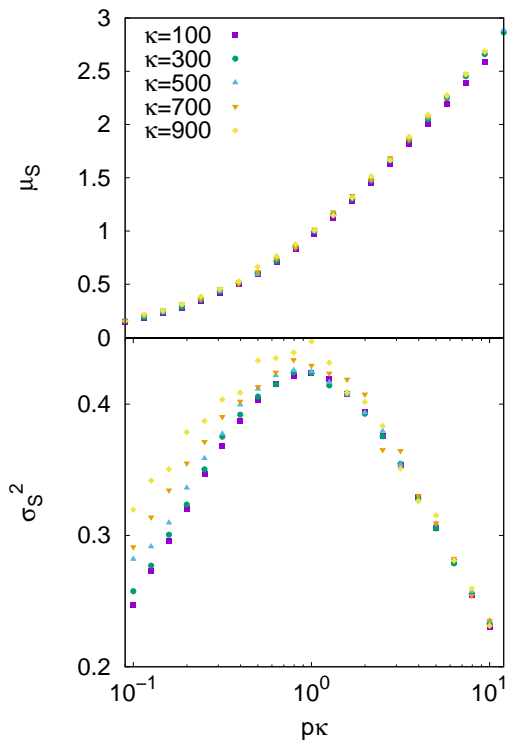}
    \caption{\label{resultados-fig3} Entropy mean $\mu_S$ (top panel) and fluctuations $\sigma_S$ (bottom panel) 
    calculated for the BMPG model for various values of $\kappa$ as a function of $p\kappa$.}
  \end{center}
\end{figure}

In order to test the existence of some form of criticality in the transition we estimate the fluctuations 
of the order parameter by computing the variance $\sigma^2_{\phi}=\avrg{\phi^2}-\avrg{\phi}^2$. 
The variance as a function of $p\kappa$ shows a peak
at $p\kappa \approx 1$ ---see inset of the bottom panel of Fig.~\ref{resultados-fig5}--- indicating that fluctuations 
reach a maximum at the inflexion point of the order parameter. 
Since all the curves collapse when plotted as a function of $p\kappa$, the magnitude of the fluctuations is independent of the size of the kernel.
Using a heuristic argument we can make an analogy with statistical mechanics by associating $p$ to the temperature---since this variable introduces disorder in the kernel---and the kernel extension $\kappa$ to the size of the system.
If we think that the probability of introducing a new element results from an activated process, then $p\propto \exp(-\Delta E/T)$, where $\Delta E$ is the activation energy. 
From this we obtain, $T \propto -1/\ln(p)$, or in a simple assumption 
$T = -1/\ln(p)$ which introduces all the temperature range.
In this scheme we can also define the susceptibility as $\chi = \frac{\kappa}{T} \sigma^2_{\phi} = -\ln(p) \kappa \sigma^2_{\phi}$.
A plot of the susceptibility $\chi$ as a function of $T = -1/\ln(p)$ is shown in Fig. \ref{resultados-fig5} (bottom panel), where it can be seen that the peak of the susceptibility  grows with the size of the kernel and is expected to diverge  at $T=0$ in the thermodynamic limit $\kappa \rightarrow \infty$, resembling the behavior of, for instance, the one dimensional Ising model.
In the top panel of Fig.~\ref{resultados-fig4} we show kernel configurations for the three main states: super-critical ($p > p_c$), critical ($p = p_c$) and sub-critical ($p<p_c$).
In the sub-critical state, the kernel is full with elements of the same class ($\phi \sim 1$), in the super-critical 
there are many elements coexisting in the kernel with similar weights and in the critical case one can distinguish the most popular element by inspection. 
Finally, a better description of the fluctuations of $\phi$ in the kernel can be obtained by calculating the distribution of the order parameter. 
We computed the distributions at the point where the fluctuations are maximum, i.e. $p=p_c=1/\kappa$, for several values of $\kappa$.
In the bottom panel of Fig.~\ref{resultados-fig4} we show these distributions corresponding to $\kappa=100, 300$ and $500$.  
The distributions do not depend on the value of $\kappa$ and  broaden around $\phi = 0.5$, further confirming the results obtained when measuring the variance. In the inset we show distributions for the sub-critical case $p<p_c$ and the super-critical case $p>p_c$ for the case of $\kappa=500$ with $p=0.1 p_c$ and $p=10 p_c$. 
Both distributions are narrow, the first one is peaked close to $1$ and the second one close to $0$, as expected.
The distribution of elements inside the kernel can be studied by computing the kernel entropy
\begin{equation}
S= - \sum_g \frac{n_{\kappa}^{(g)}}{\kappa} \ln \left( \frac{n_{\kappa}^{(g)}}{\kappa} \right),
\end{equation}
where, as before, $n_{\kappa}^{(g)}$ is the popularity of the $g$-th class of elements inside the kernel. 
In Fig.~\ref{resultados-fig3} we show the mean value of the entropy, $\mu_S$ (top panel) and its fluctuations, $\sigma_S$ (bottom panel), as a function of $p\kappa$, and for several values of $\kappa$. 
Similar to what was observed for the order parameter, all the entropy curves collapse when plotted as a function of $p \kappa$. 
The variance of the entropy behaves as the fluctuations of the order parameter, reaching a maximum value at $p \sim p_c$, also indicating the existence of a transition. 
However, it seems to increase with the size of the kernel. 
The entropy fluctuations are related to the presence of burstiness, since they imply that the rate at which an element is copied is a varying quantity.

It is interesting to note that the bursty activity of solar flares has a distribution of inter-event times with 
a power-law tail with exponent $\sim 3$, in accordance with the results of BMPG model. 
The distribution of solar flares has been satisfactorily explained using a time-dependent Poisson process that results 
from the superposition of piecewise-constant Poisson processes~\cite{Wheatland00AJ-solar-flare-mechanism}. 
Within this approach, the process is decomposed in time intervals, in which the inter-event time  is consistent with a 
constant rate Poisson process.
In analogy with solar flares, the copy rate of a given class of elements varies in the BMPG model when it is near the 
transition, as evidenced by the fluctuations of the entropy and the order parameter.
The mentioned mechanism is robust against variations of the distribution of Poisson rates. 
Hence, it can be easily adapted to explain the distribution of inter-event times of the BMGP model. 

\section{Discussion}

In this work we have studied and characterized a preferential growth stochastic model 
with a bounded memory kernel. Specifically, we have modified the Yule-Simon stochastic process by introducing a    
finite-sized memory kernel of extension $\kappa$ and called this model the Bounded Memory Preferential Growth model (BMPG model). 
We studied several statistical properties of the series of elements generated by BMPG model using numerical
simulations and standard tools from stochastic processes.

We found that the lifetime distributions of the different classes of elements in the series follow a power-law. 
We derived the master equation that rules the probability of having $n$ copies of a given 
class in the kernel at time $t$ and we found that this equation is similar to the ones 
proposed for models of species lifetime in ecologies~\cite{Pigolotti2005}, which consists in
a birth and death process. Moreover, the exponents of the obtained distributions from the BMPG model, 
$\sim 1.9$, are in the range of the reported in empirical systems~\cite{Sneppen1995}.

The BMPG model also generates elements whose popularity is distributed 
according to a power-law. This is in accordance with the highly skewed and long-tailed 
distributions found in the original Yule-Simon model and in the modified version introduced 
by Cattuto et al., which includes a long-tailed memory kernel. 
In particular, the exponent of the distribution ($3/2$) in the case of the finite-sized kernel 
can be explained  in terms of a branching process ruled by a binomial distribution, where the  
maximum number of children each node can have is equal to the size of the kernel. 
This approach works well for large enough kernels and also is able to explain the exponential 
cut-off observed in the distributions obtained with our model.

As expected, the correlations observed in the series of elements produced by BMPG model 
are short-ranged, unlike the long-range correlations observed in Cattuto's model. 
However, the correlation length in the finite-sized kernel grows nearly quadratically with 
the kernel extension.

An interesting effect observed in the presence of a finite-sized kernel---that is not observed 
neither in the Yule-Simon model nor in Cattuto's model---is that the tail of the inter-event time distributions of 
elements of the series, with a defined level of activity, decays as a power-law with exponent $\sim 3$ 
and is independent of the activity level of the pool of elements. 
This means that the generated sequences of a pool of elements show bursts of activity followed by periods 
of latency, at least for a given range of the model parameters. 
Moreover, the inter-event time distributions collapse when rescaled by the level of activity of the pool of elements 
and, in this respect, we also found that the level of activity increases with the popularity of the elements in the series. 
The burstiness parameter is greater than zero in a range of the model's parameters, reinforcing 
the evidence of the presence of burstiness in the generated series. 
The mechanism of this bursty dynamics can be associated to a superposition of
Poisson processes with a particular distribution of rates, in particular this mechanism was used to explain the
inter-event time distribution of solar flares~\cite{Wheatland00AJ-solar-flare-mechanism}
that, like in BMPG model, is distributed by a power-law with exponent $3$. 
In contrast, our previous studies of the Yule-Simon and Cattuto's models showed that the inter-event time 
distribution is well explained  by a 
Poisson process with a single rate~\cite{Schaigorodsky16PO}.

To explain the presence of the bursty dynamics in the sequences generated by the BMPG model, 
we characterized the state of  the memory kernel defining an order parameter that measures the 
occupation ratio of the kernel by the most popular class of elements and, 
also, as a measure of the distribution of classes of elements inside the kernel, we computed the kernel entropy. 
Studying the mean value and fluctuations of the order parameter and the entropy, we found that the state of the 
kernel goes through a transition from an ordered state (low values of $p$) to a disordered state (high values of $p$) 
and that there is a critical point at $p=p_c=1/\kappa$ where the fluctuation of both quantities are maximum. 
Particularly, entropy fluctuations can be closely related to the emergence of burstiness in BMPG model, 
since it implies that the rate at which an element is copied is a varying quantity. 
Moreover, since all the curves collapse when plotted as a function of $p\kappa$, the magnitude of the fluctuations 
of the order parameter and the entropy result independent of the size of the kernel.

Finally, in complementary results not shown in this paper, in which
we repeated all the analysis using an exponential short-ranged memory kernel $Q(\Delta t) = \frac{1}{\kappa_e} \exp(-\Delta t/\kappa_e)$  
with decaying rate $\kappa_e$, we found similar behaviors for the popularity and inter-event time distributions to those 
obtained with the finite-sized kernel. This suggests that these statistical properties  are independent of 
the specific functional form of the memory kernel, depending instead on how fast the distribution decays,
i.e. on the asymptotic behavior for large $\Delta t$.

\begin{acknowledgements}

This work was partially supported by grants from CONICET (PIP 112
20150 10028), and SeCyT---Universidad Nacional de C\'ordoba (Argentina).
\end{acknowledgements}


%

\end{document}